\documentclass[useAMS,usenatbib,psfig]{mn2e}

\newcommand{\mpch}{\ensuremath{h^{-1}{\rm Mpc\,}}}
\newcommand{\msun}{\ensuremath{M_{\odot}}}

\newcommand{\zetar}{${\zeta(s,q,\theta)}$}
\newcommand{\Qz}{${Q_z (s,q,\theta)\,}$}

\usepackage{psfig}
\usepackage{longtable}
\usepackage{lscape}

\title{The Three-Point Correlation Function of Luminous Red Galaxies in the Sloan Digital Sky Survey} 
\author[Kulkarni et al.]
       {Gauri V. Kulkarni,$^1$  Robert C. Nichol,$^{1,2}$ Ravi K. Sheth,$^{3}$ Hee-Jong Seo,$^4$ 
\newauthor       
       Daniel J. Eisenstein,$^{4}$ Alexander Gray$^{5}$\\
$^1$Dept. of Physics, Carnegie Mellon University, 5000 Forbes Avenue, Pittsburgh, PA 15213, USA\\
$^2$Institute of Cosmology and Gravitation, University of Portsmouth, Portsmouth, PO1 2EG, UK\\
$^3$Dept. of Physics and Astronomy, University of Pennsylvania, Philadelphia, PA 15105, USA\\
$^4$Steward Observatory, University of Arizona, 933 N. Cherry Ave.,Tucson, AZ 85121, USA\\
$^5$College of Computing, Georgia Tech,  801 Atlantic Drive, Atlanta, GA 30332, USA}

\begin{document}

\date{Accepted . Received ; in original form }

%\pubyear{2002}

\maketitle

%\label{firstpage}

\begin{abstract}

We present measurements of the redshift-space three-point
correlation function of 50,967 Luminous Red Galaxies (LRGs) 
from Data Release 3 (DR3) of the Sloan Digital Sky Survey (SDSS). 
We have studied the shape dependence of the reduced three-point 
correlation function ($Q_z(s,q,\theta)$) on three different scales, 
$s=4$, $7$ and 10 $h^{-1}{\rm Mpc}$, and over the range of $1<q<3$ and 
$0^{\circ} < \theta < 180^{\circ}$.  On small scales ($s=4\,h^{-1}{\rm Mpc}$), 
$Q_z$ is nearly constant, with little change as a function of $q$ 
and $\theta$. However, there is evidence for a shallow U--shaped 
behaviour (with $\theta$) which is expected from theoretical 
modeling of \Qz . On larger scales ($s=7$ and 10 $h^{-1}{\rm Mpc}$), the 
U--shaped anisotropy in $Q_z$ (with $\theta$) is more clearly 
detected. We compare this shape--dependence in $Q_z (s,q,\theta)$ with that seen 
in mock galaxy catalogues which were generated by populating the 
dark matter halos in large N--body simulations with mock galaxies 
using various Halo Occupation Distributions (HOD). 
We find that the combination of the observed number density of
LRGs, the (redshift--space) two--point correlation function and \Qz
provides a strong constraint on the allowed HOD parameters ($M_{min}$,
$M_1$, $\alpha$) and breaks key degeneracies between these
parameters. For example, our observed \Qz disfavors mock catalogues
that overpopulate massive dark matter halos with many LRG satellites. We also estimate the linear bias of LRGs to be $b=1.87\pm 0.07$ in excellent agreement with other measurements.
\end{abstract}

\begin{keywords}
methods: statistical -- surveys -- galaxies: statistics -- 
cosmology:  large-scale structure of universe -- cosmology: observations
\end{keywords}

\section{Introduction}
The use of correlation functions as probes of the large-scale structure 
in the Universe has a well established history in cosmology 
\citep[see][and references therein]{peb80}.  The lowest order 
correlation function, the two--point correlation function (2PCF), 
compares the number of pairs of particles (dark matter or galaxies) 
as a function of their separation, to that expected from a random 
distribution.  It is therefore, an indicator of the strength of 
clustering. Next in the hierarchy is the three--point  correlation 
function (3PCF), which compares the number of particle triplets, as a 
function of the triangle shape, to a random distribution. In recent 
years, with the advent of large galaxy catalogues like the Sloan Digital 
Sky Survey (SDSS), there has been considerable interest in accurate 
measurements of the 2PCF and 3PCF.  Combining measurements of both 
2PCF and 3PCF provides significantly improved constraints on 
cosmological parameters \citep{Sefu06}.  

Measuring the 3PCF is computationally intense as the number of 
triplets scales as $N^3$ and thus becomes prohibitive for large 
samples. In the last few years, optimal algorithms such as NPT 
\citep{Moore01, Nichol03, Gray04} and similar tree-based 
algorithms \citep{Szapudi01} have been developed, making studies 
of the higher-order clustering tractable on modern-day computational 
clusters and grids \citep{Nichol05}. We use the NPT  algorithm in this paper.

In addition, there have been recent improvements in our 
understanding of the 3PCF \citep{Scoccimarro01, Cooray02, TJ03, 
Wang04}.  These are based on the ``halo model" approach, which 
relates the galaxy 3PCF to that of the underlying dark matter.  
This approach allows measures of the 3PCF to constrain cosmology 
and galaxy formation models.  

Issues of computation and interpretation aside, the measurement 
itself has been difficult, i.e., previous measurements of the 3PCF have 
been greatly affected by the presence of rare large--scale 
structures in the data.  For example, Croton et al. (2004), 
Gazta{\~n}aga et al. (2005) and Nichol et al. (2006) have all shown 
that superclusters present in the the 2dFGRS \citep{Folk99, Coll01} and 
SDSS Data Release One (DR1) can influence the 3PCF significantly on 
large scales ($\sim 10$ \mpch).  These authors point out the need 
for samples with large enough volume to ``average out'' such 
large--scale structures.  The SDSS Luminous Red Galaxy (LRG) sample 
is well-suited for this purpose as it surveys a volume of
$\sim 1 \, {\rm Gpc^3} h^{-3}$ \citep[see Figure 1 in][]{Eis05} 
for the comparison of effective volumes for different surveys. 

This paper is organized as follows. In Section \ref{sec:fordata}, we
present the sample of LRGs used in our analysis and the measurement of
the LRG 3PCF with a full error analysis. In Section \ref{sec:Mockcat},
we discuss our method for preparing mock LRG catalogues,
while Section \ref{sec:mockpcfs} contains the measurements of 2PCF and
3PCF from these mock catalogues. We discuss our
results in Section \ref{sec:discus} and conclude in Section
\ref{sec:concl}. Throughout the paper, we have assumed a $\Lambda$CDM
cosmology with $\Omega_{\Lambda} = 0.7$, $\Omega_m = 0.3$ and $H_0 =
100\, {\rm km\, s^{-1}\, Mpc^{-1}}$ ($h=1$) unless stated otherwise.

\section{SDSS Data and 3PCF Measurement}
\label{sec:fordata}

\subsection{Data}
\label{sec:data}

The Sloan Digital Sky Survey is discussed in a series of technical
papers (Fukugita et al. 1996, Gunn et al. 1998, York et al. 2000, Hogg
et al. 2001, Stoughton et al. 2002, Strauss et al. 2002, Smith et al. 
2002, Pier et al. 2003, Blanton et al. 2003b, Ivezic et al. 2004, 
Abazajian et al. 2005, Gunn et al. 2006, Tucker et al. 2006). In
this paper, we use 50,967 spectroscopically selected LRGs as discussed
in detail by Eisenstein et al (2005) and were used to detect and
measure the Baryon Acoustic Oscillation peak in the large-scale
2PCF. The details of the LRG selection algorithm are discussed in
Eisenstein et al. (2001).  Briefly, our sample is based on the SDSS
Data Release 3 (DR3) and spans a redshift range of $0.15 < z < 0.55$
with a g-band absolute magnitude range of $-23.2 < M_g < -21.2$. The
comoving density is $9.7 \times 10^{-5} {\rm Mpc^3} h^{-3}$, which is
approximately constant upto a redshift of $z = 0.36$ and drops
thereafter \citep{Eis05}. For the calculations of the correlation
functions \citep[see][] {SS98}, we use exactly the same random
catalogues as Eisenstein et al. (2005), which contain $16$ times the
number of galaxies as in the real data and have the same selection
function as the real data.

\subsection{Results of 3PCF measurement}
\label{sec:data3pcf}

In order to study the dependence of 3PCF on triangle configuration,
one needs to parametrize the shape of the triangle. If $s_{12}$, $s_{23}$,
$s_{31}$ are the lengths of the three sides of a triangle, then a
commonly used parametrization is given as $(s,q,\theta)$ with $s =
s_{12}, \quad q = s_{23}/ s_{12}$ and $\theta$ the angle between
$s_{12}$ and $s_{23}$. Then we can define the reduced 3PCF (\Qz), which is
the ratio of the 3PCF (\zetar), to the sum of the products of the
2PCFs for the three sides (see Groth \& Peebles 1977), or

\begin{equation}
Q_z = \frac{\zeta(s,q,\theta)}{\xi(s_{12})\xi(s_{23}) + \xi(s_{23})\xi(s_{31}) + \xi(s_{31})\xi(s_{12}) } .
\label{qz}
\end{equation}

\noindent This is also known as the ``hierarchical ansatz"
\citep{peb80} and the subscript $z$ denotes measurements made in
redshift--space.

\begin{figure}
\centerline{\psfig{file=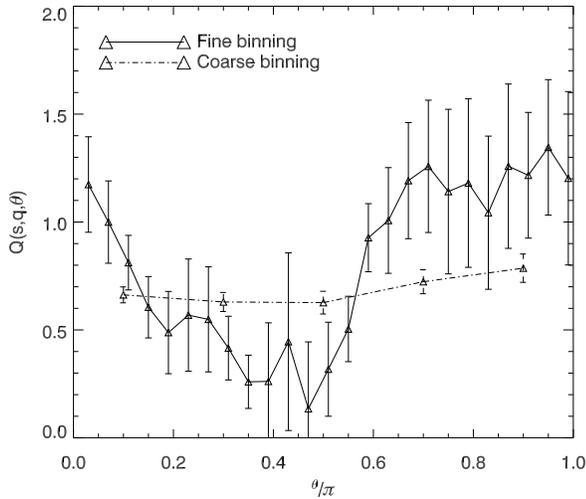,width=3.5in}}
  \caption{The effect of binning on the shape of the observed \Qz. We
  compare here our measurement for the reduced 3PCF for the triangle
  configuration of s = 10 \mpch and $q = 2$. Also shown is the
  coarse binning scheme of Nichol et al. (2006) for the same triangle
  configuration. Errors on both measurements are from jack-knife
  resampling (see text for explanation).  The coarser (wider) bins
  tend to ``smooth out'' features in the \Qz as proposed by GS05. 
  \label{bincomp}
  }
\end{figure}

One recent issue discussed by Gazta{\~na}ga \& Scoccimarro (2005; GS05
henceforth) is the effect of binning resolution on the shape of the
observed \Qz which potentially hinders our ability to measure the
characteristic U-shaped anisotropy (between ``open'' and ``collapsed''
triangle configurations) witnessed in N-body simulations (GS05,
Fosalba, Pan \& Szapudi 2005).  The first measurements of the 3PCF
from the 2dFGRS \citep[e.g.][]{GA05} and SDSS \citep[e.g.][]{Nichol06}
show evidence for this expected U-shaped dependence, but it is not as
strong as expected from simulations. Therefore, we measure our \Qz in
as narrow bins of $s, q$ and $\theta$ as possible. We use $\Delta s =
0.2$ \mpch and $\Delta q = 0.2$ for $s$ and $q$ respectively, while we
use $\Delta \theta = \pi/50$, giving potentially 50 bins in $\theta$
(we actually only measure the alternate bins in $\theta$, giving 25 in
total). Making the $\theta$ bins any smaller than this value would
lead to very small triangle counts in each bin (i.e. $< 10$ triplets
in some bins), which would then require larger bins in $s$ and
$q$. This binning scheme represents the narrowest set of bins we can
construct for our sample of LRGs. However, as demonstrated in Figure
\ref{bincomp}, this binning scheme is much better than previous
measurements and does resolve the shape--dependence of $Q_z$.

\begin{figure}
\centerline{\psfig{file=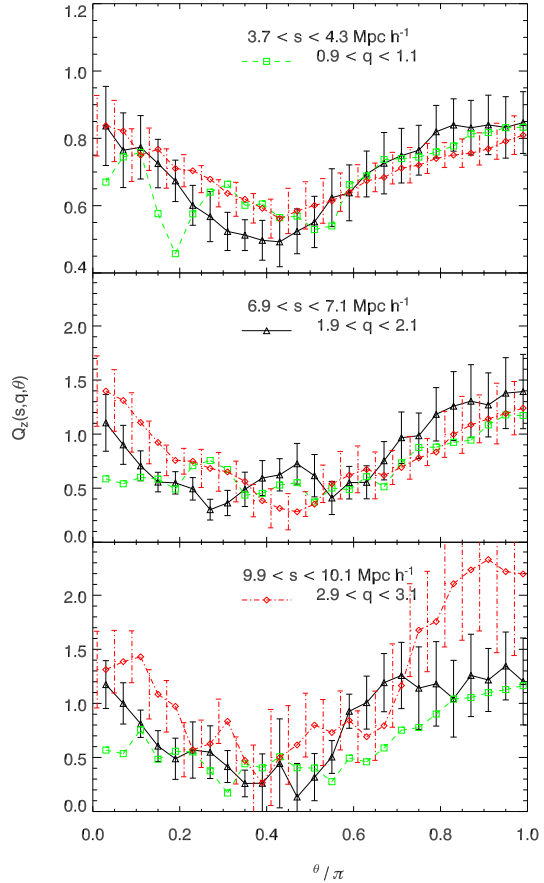,width=3.25in}}
%  \epsscale{0.85}
 % \plotone{lrg3pcf.pub.ps}
  \caption{Our measurements of \Qz for three different ranges of triangle scale ($s$; see label in each panel) 
  and different $q$ values (see all panels for definitions). The dependence of
  \Qz on the $q$ parameter does not appear to be strong. We do see the
  gradual emergence of the expected U-shaped behaviour of \Qz going
  from $s=4$ to $s=10$ \mpch. We do not plot the errorbars for $0.9 <
  q < 1.1$ to avoid overcrowding while errorbars for $2.9 < q < 3.1$
  are plotted with an artifical offset of -0.02. Note that the y--axis in the top panel covers a small range of \Qz values than the other two panels. Also, the $s$ ranges shown only apply to the panels they appear in, while the $q$ symbol definitions apply to all three panels}
  \label{lrg3pcf}
\end{figure}

\begin{figure}
\centerline{\psfig{file=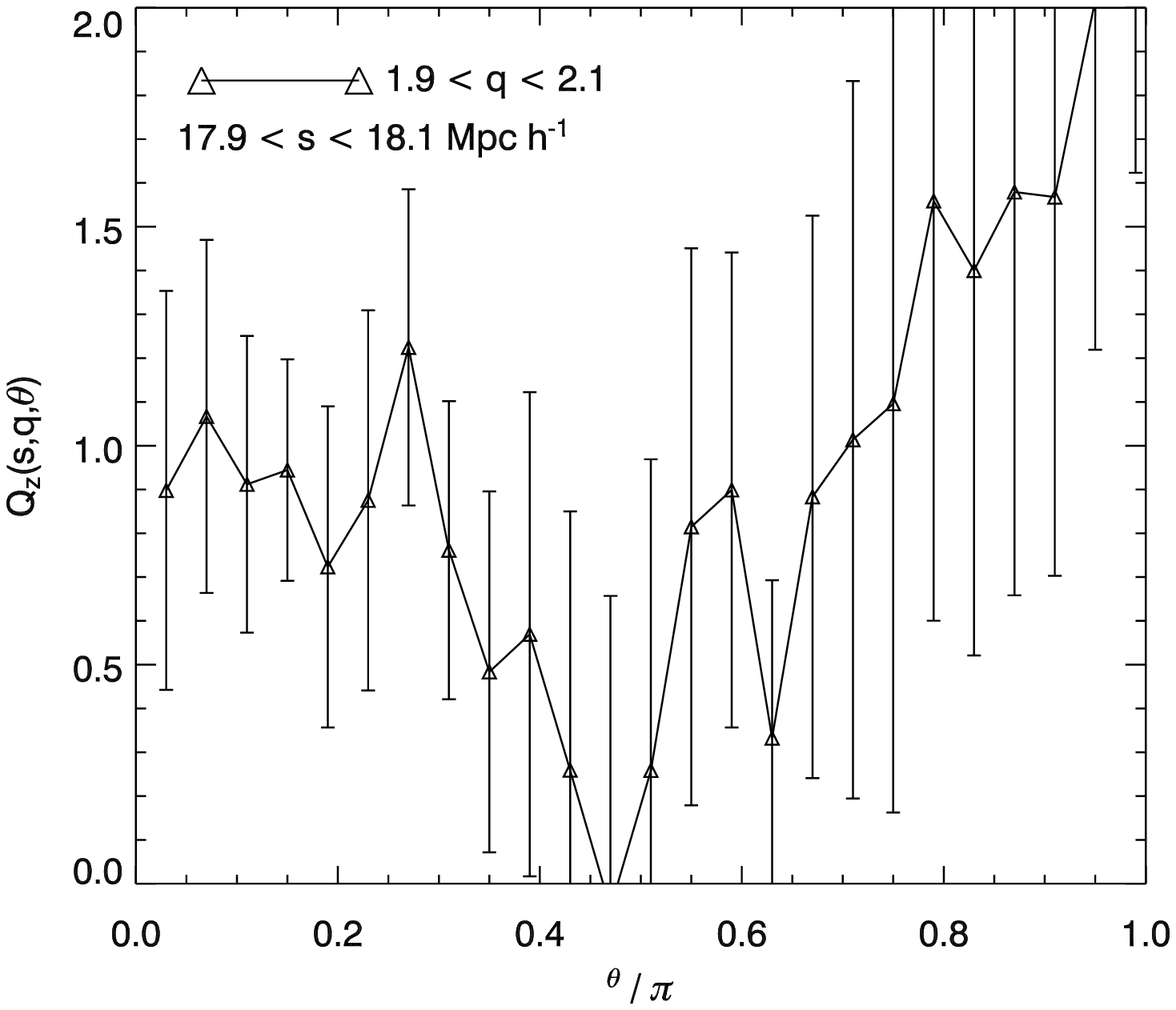,width=3.25in}}
  %\plotone{s9err.ps}
  %\plotone{compwidth.ps}
  \caption{The measurement of \Qz on large scales, i.e., $s=18$ \mpch.
  The errors on this measurement are large, making it hard to
  unambiguously detecting any shape--dependence.  The solid black line
  (with triangle symbols) is the measurement of \Qz with the same bins
  as used in Figure \ref{lrg3pcf}.}
  \label{s9err}
\end{figure}

In Figure \ref{lrg3pcf}, we show our measurement of the
redshift--space reduced 3PCF, \Qz, on scales of $s=4$, 7 and 10
\mpch. We also present these data in Tables A1, A2 and A3 in the
Appendix. These scales closely match the $s$ scales used in Figure 2
of GS05 (3, 6 and 12 \mpch\, respectively in their
figure). Measurement of the 3PCF become unreliable on scales less than
4 \mpch\, because of the small number of triplets in each bin, i.e.,
we have $<10$ triangles. On $s$ scales greater than 10 \mpch, our
measurement of the 3PCF from this LRG sample become noisy, e.g., at
$s=18$ \mpch\, (a scale used in GS05), it is difficult to detect the
shape--dependence in the \Qz given the large error bars (see Figure
\ref{s9err}).

\subsection{Determination of Errors}
\label{sec:error3pcf}

\begin{figure}
  \centerline{\psfig{file=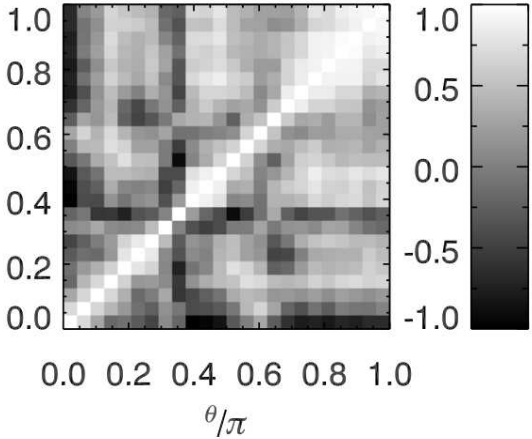,width=3.0in}}
  \caption{We present here the normalized covariance matrix for the 3PCF for the
  triangle configuration of $9.9 < s < 10.1$ \mpch\, with $1.9 < q <
  2.1$. See Figure \ref{s7diff} for further explanation.}
  \label{covar}
\end{figure}

The error bars on both the 2PCF and 3PCF were estimated using the
jack-knife resampling technique \citep{Scran02, Zehavi05}. We spilt
our sample into 11 (almost) equal area subsets, and then by omitting
each of these subsets one-by-one, we repeat the measurement of $Q_z$
eleven times to compute the r.m.s. variation between these
measurements. This process provides an estimate of the covariance
matrix, with the diagonal element of these matrix shown as error bars
in our figures. We present in Figure \ref{covar} an example of one of
our covariance matrices.

\begin{figure*}
  \centerline{\psfig{file=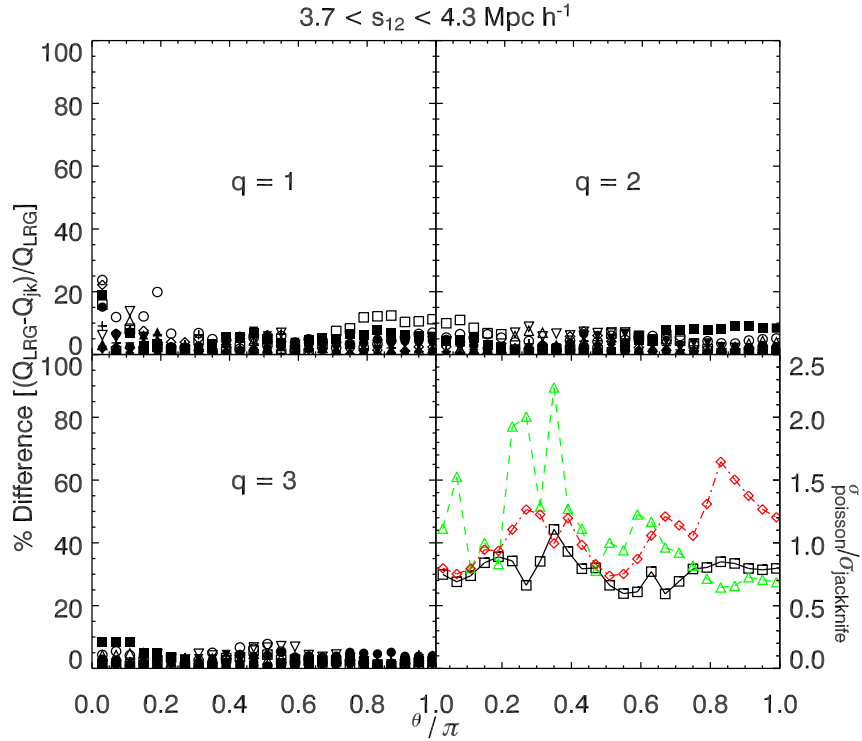,width=4.5in}}
  %\epsscale{0.8}
 % \plottwo{s1diff_pub.ps}{s4diff_pub.ps}
  %\plotone{s1diff_pub.ps}
  \caption{The absolute percentage difference between the 3PCFs of the
  11 jack-knife datasets discussed in the text and the full dataset
  for the $s=4$ \mpch scale (first three panels). No single jack-knife
  estimation dominates the difference between these measurements, thus
  indicating the LRG sample is a fair sample of the Universe on this
  scale. Last panels (lower right) shows the comparison of the errors
  with the expected Poisson errors.  The green triangles are for q =
  1, black squares are for q = 2 and red diamonds are for q = 3. }
  \label{s1diff}
\end{figure*}

\begin{figure*}
  \centerline{\psfig{file=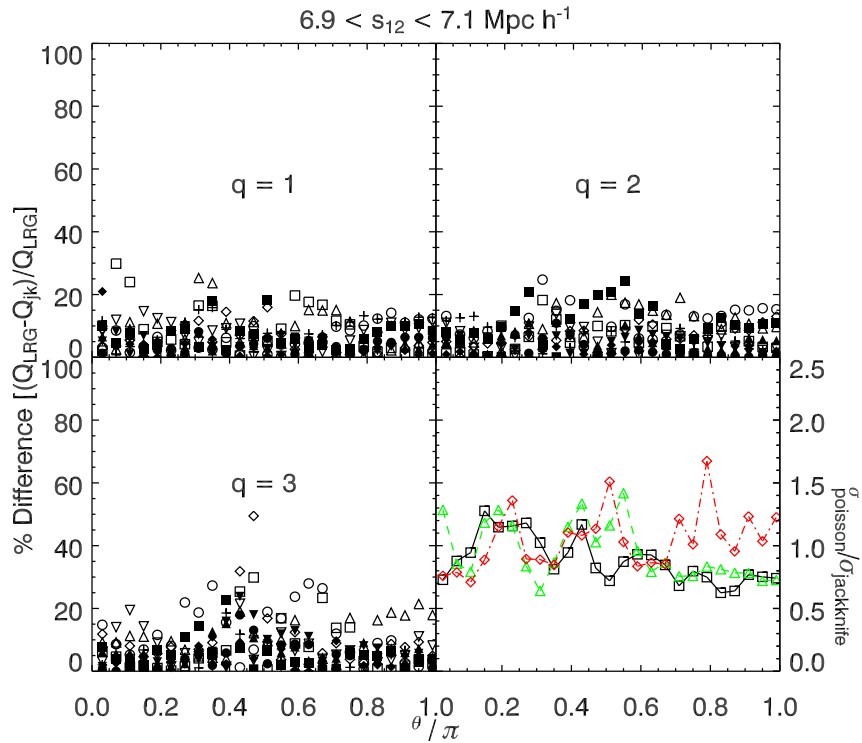,width=4.5in}}
  %\epsscale{0.8}
  %\plotone{s4diff_pub.ps}
  \caption{The same as Figure \ref{s1diff}, but for the scale $s=7$
  \mpch scale.}
  \label{s4diff}
\end{figure*}

\begin{figure*}
  \centerline{\psfig{file=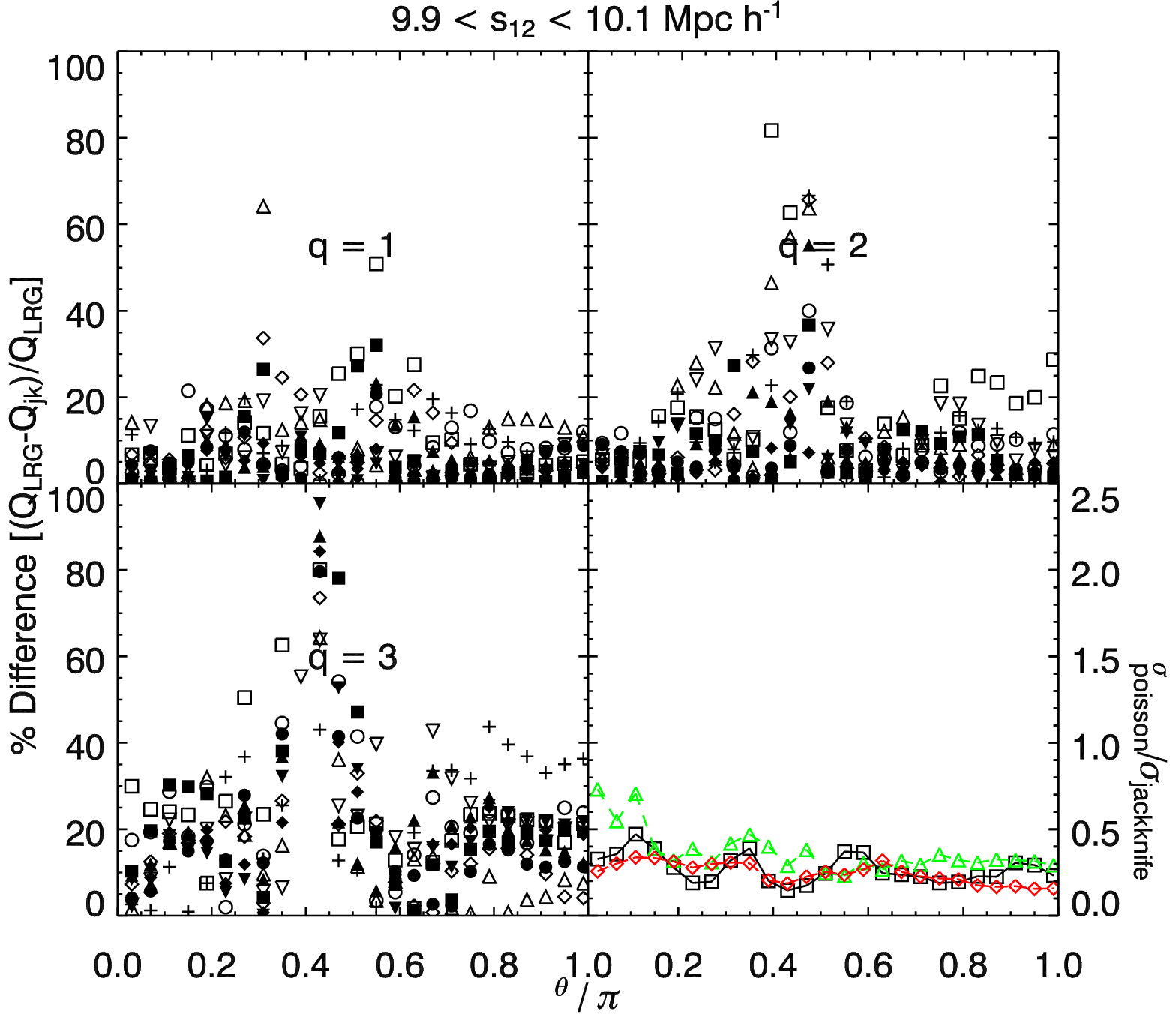,width=4.5in}}
  %\epsscale{0.8}
  %\plotone{s7diff_pub.ps}
  \caption{The same as Figure \ref{s1diff} but for the $s=10$ \mpch
  scale. However, on this scale, although no single jack--knife region
  dominates, there is considerably larger variations between the
  different regions. Also, the jack-knife errors are now significantly
  larger than the expected Poisson errors (as indicated in the lower
  right panel), which indicates the errors are correlated on these
  larger scales as expected because of large-scale structures in the
  Universe.}
  \label{s7diff}
\end{figure*}

One of the advantages of using the jack-knife error technique is
demonstrated in Nichol et al. (2006) who used the different \Qz
measurements for the different jack-knifes samples to investigate the
influence of large--scale structures (e.g. isolated in a single
sub--region) on the measurement of the entire sample of galaxies. In
Figures \ref{s1diff}, \ref{s4diff} and \ref{s7diff} we present the
individual measurements of $Q_z$ for the eleven jack--knife samples
for the three scales, i.e., $s=4$, 7 and 10 \mpch. The first three
panels of Figures \ref{s1diff}, \ref{s4diff} and \ref{s7diff} show the
absolute value of the percentage difference between the individual
eleven jack-knife measurements compared to the measurement from the
whole LRG sample. These panels demonstrate that there is no single
``rogue'' sub--region that dominate the error on the 3PCF, unlike the
findings of Nichol et al. (2006) that found the 3PCF of the SDSS main
galaxies was dominated by the presence of the ``Sloan Great Wall''.

The last (lower right) panel of Figures \ref{s1diff}, \ref{s4diff} and
\ref{s7diff} compare our estimation of the errors on \Qz for the whole
LRG sample with the expected Poisson error for the number of triplets
in each bin. For the 4 and 7 \mpch scales, the Poission errors are
approximately the same as the jack--knife errors indicating that on
these scales the main source of error is simply
shot--noise. Therefore, bigger samples of LRG galaxies will improve
the measurement of the 3PCF on these small scale.  However, in Figure
\ref{s7diff} (right panel), we see a different behavior, with the
jack--knife errors being three times larger than the Poission
errors. This demonstrates that on these larger scale the correlation
function bins are highly correlated by the large-scale structure in
the Universe. The stability of the jack--knife errors on these large
scales ($s=10$ \mpch) also confirms that the jack--knife technique has
captured such correlations between the bins and is a better measure of
the true error on these scales.

\section{Correlation Functions of Mock Catalogues}
\label{sec:formocks}

%% End of Section.

\subsection{The Halo Model}
We use mock catalogs based on the halo model to interpret our 
measurements of the LRG 3PCF.  Briefly, let $p(N|M)$ denote 
the probability that a halo of mass $M$ contains $N$ LRGs.  
The first moment of this probability distribution, 
$\langle N|M\rangle$, gives the mean number of LRGs in a halo of 
mass $M$; it is sometimes called the Halo Occupation Distribution
(HOD).  We use the following parameterized form for this relation:  
\begin{eqnarray}
 \left < N_{cent}(M)\right > & = & \mbox{exp}\left(- \frac { M_{min} }{M} 
 \right),
 \label{Ncent} \\
 \left < N_{sat}(M)\right > & = & \mbox{exp}\left(- \frac { M_{min} }{M}
 \right) \left( \frac{M}{M_1} \right) ^ \alpha ,
 \label{Nsat} \\
 \left < N(M) \right> & = & \left<N_{cent}(M)\right> + \left<N_{sat}(M)
 \right> .
 \label{NM}
\end{eqnarray}
The terms central and satellite have the following meanings.  
In halos which host more than one LRG, the first LRG is placed 
at the halo center and is called the central galaxy; the others 
are distributed around this center and are called satellites.  
The expressions above show that there is assumed to be a reasonably 
sharp mass threshold $M_{min}$ below which halos are increasingly 
(exponentially) unlikely to host even one (central) LRG.  
More massive halos which host a central LRG may also host 
satellites, the typical number of which is assumed to scale as 
a power law in halo mass:  $M_1$ denotes the mass required to 
host at least one satellite, and the slope of the power law, 
$\alpha$, describes how quickly the mean number of satellites 
increases with mass.  
Zehavi et al. (2005) have shown that this parametrization, 
with $M_1\approx 23M_{min}$ and $\alpha\approx 1$ 
provides a good desription of the HOD for galaxies above some 
threshold luminosity in the SDSS Main Galaxy Sample ($M_{min}$ 
increases with increasing threshold luminosity).

The expressions above specify only the mean of $p(N|M)$.  
When inserted into the halo model, this is only sufficient to 
estimate the mean number density of LRGs.  To predict the 2PCF, 
a model for the second moment of this distribution is required; 
the third moment of $p(N|M)$ is required for the 3PCF.  
We specify the entire hierarchy of correlation functions by 
assuming that the number of satellites in a halo which contains 
a central LRG is drawn from a Poisson distribution with mean 
$N_{sat}(M)$.  This Poisson satellite assumption is motivated 
by results in Kravtsov et al. (2004), and was used by 
Zehavi et al. (2005) in their study of the luminosity 
dependence of the 2PCF in the SDSS Main Galaxy Sample.

\subsection{Mock Catalogues}
\label{sec:Mockcat}

We begin with dark matter halo catalogues generated from a set of six cosmological N-body
simulations with a box size of $512^3{\rm (h^{-1}Mpc)^3}$, containing $256^3$
particles of mass 8.28 $\times 10^{11}$ \msun. The N-body simulations are generated
using the Hydra code (Couchman et al. 1995) in collisionless
particle-particle-particle-mesh (P$^3$M) mode with $256^3$ force grids and a
Plummer softening length of 0.2\mpch (see Seo \& Eisenstein 2005 for more
details about these simulations). The cosmological parameters used to
generate the simulation are $\Omega_m = 0.27$, $\Omega_{\Lambda} = 0.73$, $h = 0.72$, 
$\sigma_8 =0.9$ and $n=0.99$, in agreement with the WMAP1 
best--fit values. Halos in these simulations were
detected using the friends-of-friends algorithm (Davis et al. 1985) with
a linking length of 0.6\mpch and a minimum group multiplicity of 20
particles. We note that these simulations do not use the more conventional $b=0.2$ linking length and
therefore, our halo masses will exceed the halo masses from more conventional simulations
(e.g. which are closer to $M_{200}$). We have checked that our mass functions are approximately the same as the Sheth \& Tormen (1999) $z=0$ mass function but with all the masses scaled by 1.5.  Our mass functions can therefore be thought of as representing the expected mass functions in the near future and, at a fixed number density, we would not expect the HOD for these massive LRGs to have evolve significantly. This should make comparisons with other HODs in the literature easier, but such issues should be taken into account when performing detailed comparisons with our best--fit $M_{min}$ and $M_1$ values, although $\alpha$ should not be different\footnote{To facilitate a detailed comparison with our HOD results, the mass functions of our simulations are available on request.}.

Initially, we use a halo catalogue generated from one of the six
simulations, henceforth called the \it parent \rm simulation, and then use 
the halo model to populate the dark matter 
halos in the simulations with central and satellite LRGs.  The halo model has three 
free parameters, and so we construct mock catalogs which 
have a range of $M_{min}, M_1$ and $\alpha$.  
Specifically, $M_{min}$ spans the range from the minimum 
halo mass in the simulation (called $M_{min}^{sim}$) 
to the maximum halo mass (called $M_{max}^{sim} $) in the 
simulation using integer steps of the value of $1 \times 10^{13}
M_{\odot} $. By varying $M_1$ from $M_{min}$ to $M_{max}^{sim}$ in the
steps of $M_{min}^{sim}$, we also find a range of $M_1$ values for each
$M_{min}$. As $N_{sat}$ in Eqns 2 \& 3 changes slowly over this
range, we sample it in logarithmic steps of 0.1.  
Using the value of $M_{min}^{sim}$ to increment $M_1$ assures we 
have a reasonably fine grid of models to test against the 
observations.  The slope $\alpha$ was allowed to vary from 0.9 
(the value obtained by Zehavi et al. (2005) for their faintest galaxy 
sample) to 4 in increments of 0.1.

\begin{figure}
  \centerline{\psfig{file=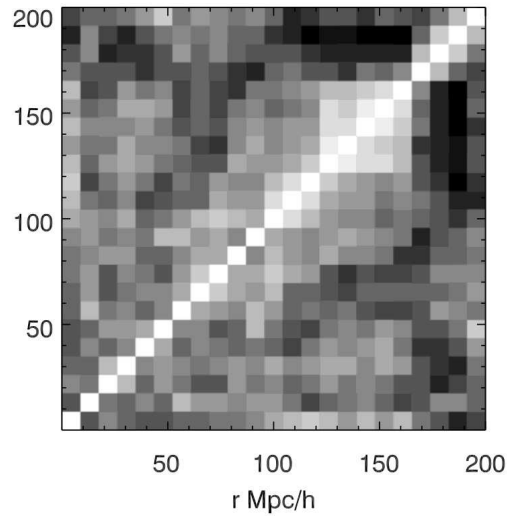,width=3.0in}}
  \caption{Covariance matrix for 2PCF measurement of the LRGs. 
  The colour scheme is same as in figure \ref{covar}.}
  \label{covar2pt}
\end{figure}

\begin{figure}
  \centerline{\psfig{file=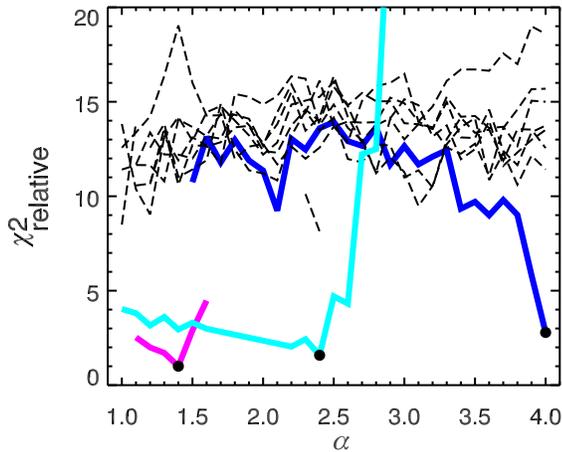,width=3.5in}}
  %\plotone{chi2alphaall_pub.ps}
  \caption{We present here slices in $\chi^2$--space (we have
  normalized these curves by the minimum $\chi^2$ of all the 354
  mocks). Each curve corresponds to a unique set of $M_{min}$ and
  $M_1$ combinations, as a function of $\alpha$. The thick, coloured
  curves are the three preferred solutions discussed in the text. The
  solid black circles on each of the solid curves corresponds to the
  solution chosen from that group for the analysis in this paper. We
  note that some of the curves do not span a full range of $\alpha$
  values because these combinations of the three HOD parameters have
  been excluded by our initial number density constraint, e.g., the
  magenta curve. Also, curves with parameters that yield (normalized)
  $\chi^2 > 60$ are not plotted here.}
  \label{chi2alpha}
\end{figure}

We compute the mean number of galaxies associated with each HOD 
and reject those cases in which the associated number density 
differs by more than 5\% from the observed mean density of LRGs 
in our sample (this latter was estimated from a volume--limited 
subsample).  Eisenstein et al. (2005) show that the comoving density 
is constant for the redshift range $0.16 < z < 0.36$.  
Only 354 different combinations of HOD parameters survive this
test.

For each of the 354 allowed set of HOD parameters, we populate the
dark matter halos in the simulations as follows. 
We bin all halos into 300 equal mass bins between $M_{min}^{sim}$
and $M_{max}^{sim}$. If $N_i$ denotes the number of halos in 
the $i$th bin, then we choose $N_i\mbox{ exp}-(M_{min}/M_i)$
halos at random from $N_i$ and assign them a central LRG. 
The central LRG is assigned the same spatial coordinates as the 
center of mass of its host halo.  
For halos which host a central LRG, the number of satellite LRGs 
was drawn from a Poisson distribution with mean given by 
Eqn \ref{Nsat}.  The positions of these satellite LRGs were 
assigned using the NFW (Navarro, Frenk \& White 1997) density profile.
% with a
%halo concentration parameter ($c(M)$) following Bullock et al. (2001),
%corrected to be consistent with the definition of halos as overdensity
%200 times critical, i.e., 
%
%\begin{equation}
%c(M) = 11\left( \frac{ \left( \frac{M}{M_*} \right) ^ {-0.13} }{1+z}
%\right)
%\end{equation}
%
%\noindent where $M_*$ is the nonlinear mass at z=0. 
Since our measurements are in redshift space, we must also model 
the peculiar velocities of our mock LRGs.  
We work in the ``distant observer approximation", meaning that 
redshift space effects are incorporated by adding peculiar velocities 
to the $z$-components of the position vectors of the mock galaxies. 
This was done as follows.  The central LRG in a halo is assumed to 
have the same velocity vector as its host, so redshift space 
distortions are due to the $z$-component of the halo's velocity 
vector, $v_z$.  The satellites are assumed to be in approximately 
virial equilibrium; to model this, the velocity of a satellite is 
given by adding a random Gaussian number with rms 
$\sigma_z \propto M_{halo} ^{1/3}$ to $v_z$ before combining 
with the $z$-component of the spatial position (Sheth \& Diaferio 2001).  
We populate the halos of the other five simulations using the set of allowed HOD parameters derived from the {\it parent} simulation (only the {\it parent} simulation satisfies the number density constraint discussed above, while the other five mocks can violate this constraint).  Thus, for each 
allowed set of HOD parameters, we have six mock catalogues all 
in redshift--space. 

\begin{figure*}
    \centerline{\psfig{file=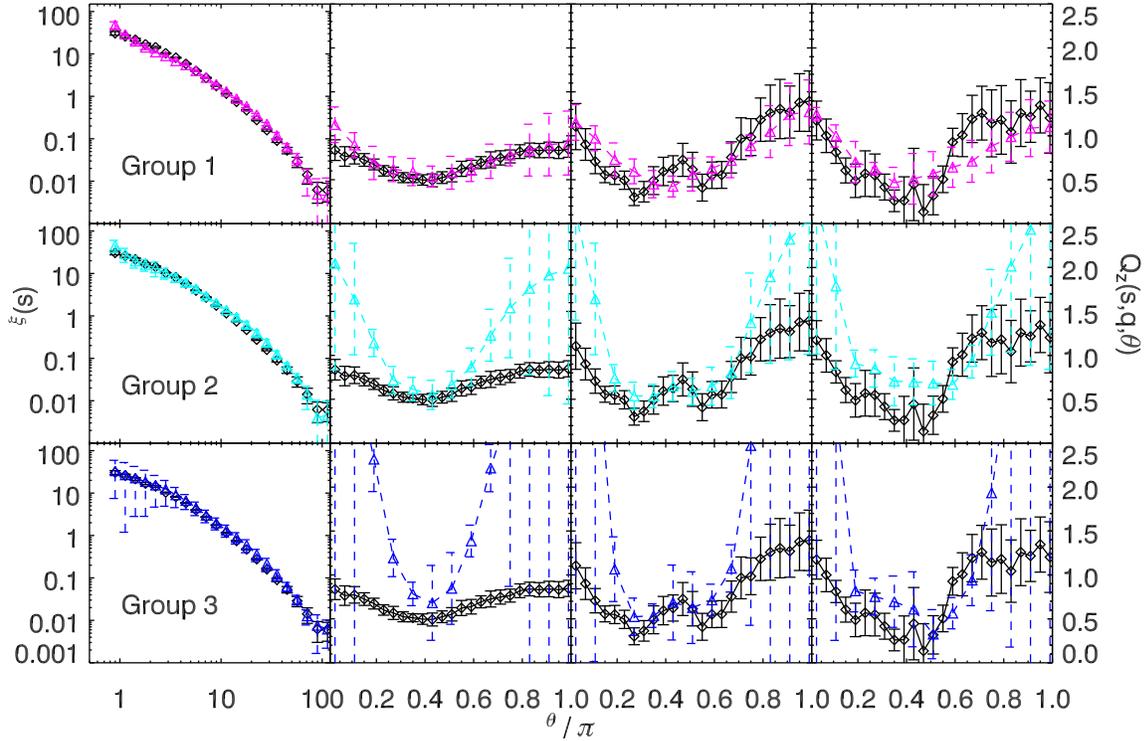,angle=90,width=6in}}
    \caption{We present the 2PCF and 3PCF of the mock LRG catalogues
    from three ``best'' mocks discussed in the text, i.e., the three
    mock catalogues with the lowest $\chi^2$.  The colour scheme
    follows the same one used in Figure \ref{chi2alpha}. For each row,
    from left to right, we show the measurements of the 2PCF (both
    mock and real data), the 3PCF at the scale $3.7 < s < 4.3$ \mpch,
    the 3PCF at the scale $6.9 < s < 7.1$ \mpch and the 3PCF at the
    scale $9.9 < s < 10.1$ \mpch. For the 3PCF, we only show the $1.9
    < q < 2.1$ bin. The measurements for the mocks are the mean and
    variance over the 6 simulations available to us. In each group, we
    see excellent agreement between the 2PCF of the mocks and real
    data. However, 3PCFs behave very differently.}
    \label{allpcf}
\end{figure*}

\subsection{Testing the Mocks}
\label{sec:mockpcfs}

In this section, we compare our mock catalogues with the real SDSS
data. Using the same initial simulation as discussed in Section
\ref{sec:Mockcat}, we measure the 2PCF for each of the 354 allowed
mocks.  We then define a $\chi^2$ from the difference between the 
2PCF measured in the mock and the real data, over the range 
$0.89 < s < 57$ \mpch\ (corresponding to 18 bins in separation). 
For this comparison, we use the full covariance matrix obtained 
for the real data (Figure \ref{covar2pt}) using 20 jack-knife 
sub--samples (instead of 11). 
We have used more jack-knife regions here than for the 3PCF
above (see Section \ref{sec:data3pcf}), as it provides a more 
stable inversion of the covariance matrix on small scales 
($0.89 < s < 9 \,\mpch$).

Figure \ref{chi2alpha} shows individual slices through the
(normalised) $\chi^2$ surface as defined by the $M_{min}$ and $M_1$
HOD parameters for the 354 allowed mock catalogues. As can be seen,
there are 3 distinct curves (or ``valleys'') in this 3D parameter space of $M_{min}$, $M_1$ and $\alpha$
that possess the three lowest minima in the $\chi^2$ surface.  These
three curves are highlighted (in colour) in Figure \ref{chi2alpha} and
have the values of $M_{min} = 7.66 \times 10^{13} \msun, \mbox{ } M_1
= 4.7 \times 10^{14}$ (magenta); $M_{min} = 7.66 \times 10^{13}
\msun, \mbox{ } M_1 = 7.6 \times 10^{14} \msun$ (cyan); and
$M_{min} = 6.66 \times 10^{13} \msun, \mbox{ } M_1 = 1.3 \times
10^{15} \msun$ (blue). In Figure \ref{allpcf}, we show the mean and
rms variation (from the six mocks) for both the 2PCF and 3PCF for
these three sets of HOD parameters with the lowest $\chi^2$ values
shown in Figure \ref{chi2alpha}. Table \ref{HODtable} provides the best fit HOD parameters.

\section{Discussion}
\label{sec:discus}

In Figure \ref{lrg3pcf}, we show the shape dependence of the reduced
3PCF (\Qz) for the Luminous Red Galaxy sample presented in Eisenstein
et al.  (2005). On small scales ($3.7<s<4.3\mpch$), our \Qz is nearly
constant for all values of $q$ (consistent with the original findings of Groth \& Peebles 1977). There
is however clear evidence for a shallow U--shaped anisotropy between
``open'' triangles (defined to be triangle configurations at the ends
of the $\theta$ range) and ``collapsed'' trianges (configurations with
intermediate $\theta$ values). The absence of a strong U-shape,
especially at the ends of the $\theta$ range, suggests the lack of
strong ``Fingers--of--God'' (FOG) for these LRGs, which is to be
expected as many halos only possess one, or a few, LRGs. For example,
approximately 50\% of cluster-size halos in our mocks (masses of
$>10^{14}$ \msun) only have a single LRG. This can be contrasted to the sharp U--shaped behaviour
(at $\theta\simeq0$ and $\theta\simeq180$ degrees) on small scales seen
in the dark matter simulations of GS05.

On larger scales ($s=7$ and $s=10$ \mpch), the broad U--shaped
anisotropy of the reduced 3PCF becomes more pronounced, due to the
emergence of the large--scale filamentary structure in the
Universe. This is now in qualitative agreement with earlier results (Frieman \& Gaztanaga 1999) and expectations from
N--body simulations (but still less than predicted, on these scales,
by GS05). A more detailed comparsion of the observed 3PCFs and dark
matter simulations will be required and is beyond the scope of this
paper.

%% Added for the referee
%%
Throughout this paper, we have used the same 3PCF parametrization of the triangle configurations as GS05 ($s$,$q$,$\theta$) to allow for easy comparison with their work and other recent measurements of the 3PCF (Gaztanaga et al. 2005; Nichol et al. 2006). However, we note that this parametrization can lead to some triangles being represented more than once in different bins. We have not corrected for this effect here but stress that our jack--knife errors will include any extra correlations due to this effect. This may explain some of the significant off-diagonal elements in the covariance matrices (Figure 4). We note that our mock catalogues have been analysed in the exact same way as the real data and therefore, also include such issues. We also use the narrowest bins possible in $s$,$q$ and $\theta$, which will minimize this effect and keep the bins as independent as possible. 
%%%

Instead, we have used the halo model to understand the LRG 3PCF in
redshift--space.  We have used dark matter halos from large N--body
simulations to create mock galaxy catalogues which are tested against
the real data using both their mean number density of galaxies and
2PCF. We find that our mock catalogues fall into three distinct
groupings (which possess the lowest $\chi^2$ fits to the real data) as
defined by their HOD parameters ($M_{min}$, $M_1$, $\alpha$).  In
reality however, our 2PCF fits (in Section \ref{sec:mockpcfs}), are
insensitive to $M_{min}$ (the minimum mass of a halo to hold a central
galaxy) because it is initially constrained by our requirement on the
number density of LRGs. Therefore, only $M_1$ and $\alpha$ are allowed
to change and as demonstrated in Figure \ref{allpcf} (first column)
there are multiple combinations of these two parameters that give good
fits to the 2PCF. Figure \ref{chi2alpha} clearly shows the degeneracy
between $M_1$ and $\alpha$. However, Figure \ref{allpcf} demonstrates
that the 3PCF can break this degeneracy between these parameters as
the \Qz from the mocks changes significantly as $M_1$ and $\alpha$ are
varied (while the 2PCF remain almost identical).

\begin{figure}
  \centerline{\psfig{file=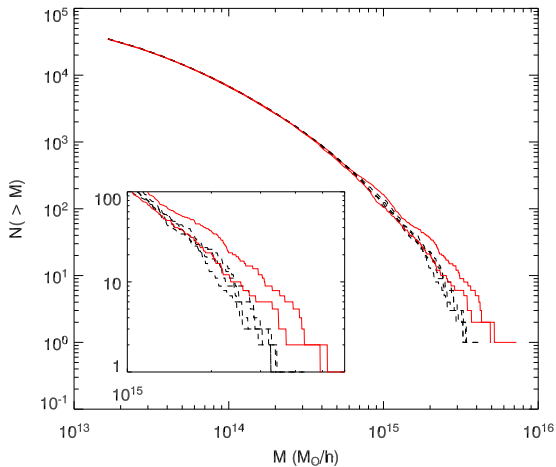,width=3.25in}}
  %\plotone{massfun_pub.ps}
  \caption{The number of halos, greater than a given mass $M$, for six 
  simulations used in this paper. The functions shown in red are the two 
  simulations excluded in Figure \ref{eallpcf} and discussed in the text.}
  \label{massfn}
\end{figure}

\begin{figure*}
    \centering{\psfig{file=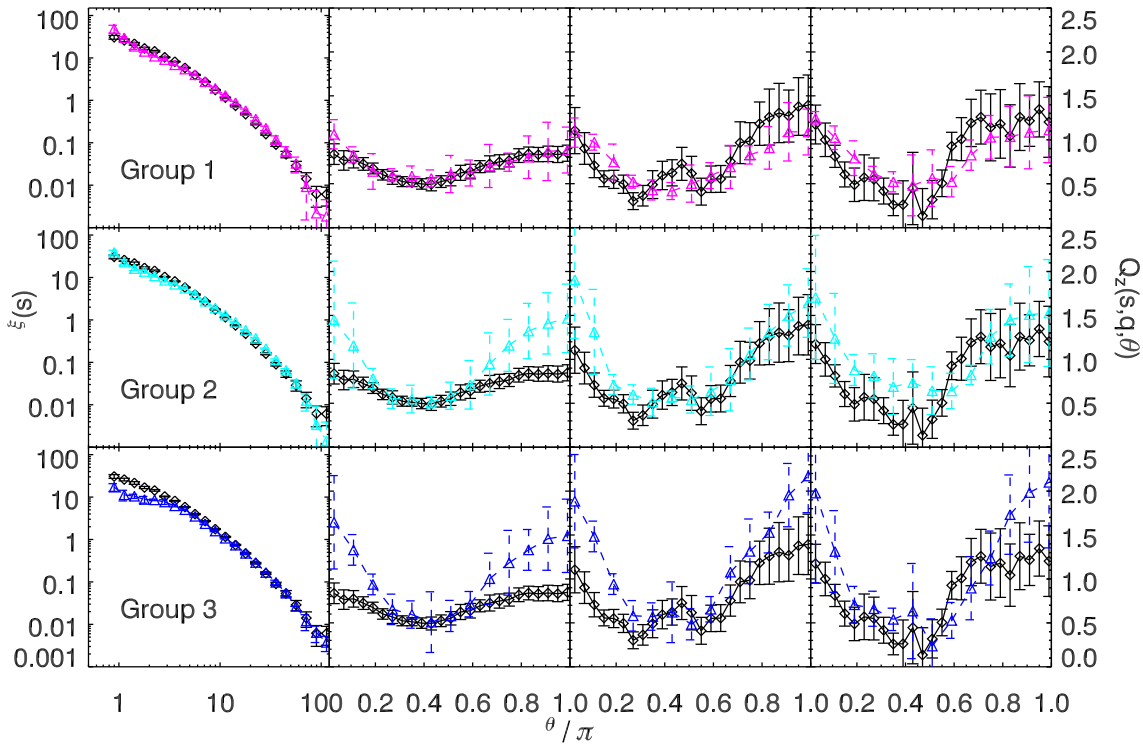,angle=90,width=6in}}
    \caption{The same as Figure \ref{allpcf}, but with \it sim1 \rm
    and \it sim2 \rm excluded (shown in red in Figure
    \ref{massfn}). Even after omitting these two simulations, the 2PCF
    of the mocks and data are still in good agreement. For the 3PCF,
    we still witness more anisotropy in the mocks for the bottom two
    mocks than observed in the data. }
    \label{eallpcf}
\end{figure*}

It is worth discussing here the size of the error bars on our \Qz
measurements for the mocks (i.e., the variance between the 6 mock
catalogues used herein) shown in Figure \ref{allpcf}. In particular,
the error bars for Groups 2 \& 3 in Figure \ref{allpcf} are
significantly larger than those for Group 1 and are caused by two of
the six dark matter simulations having significantly more
massive halos than the other four. This is illustrated in Figure
\ref{massfn}. We denote these two ``rogue'' simulations as \it sim1 \rm
and \it sim2\rm, and if we omit these two simulations when computing
the mean and variance, then our fits to the 2PCFs and 3PCFs are
significantly better with smaller errors, see Figure \ref{eallpcf}
(columns 2, 3 and 4).

\begin{figure}
  \centerline{\psfig{file=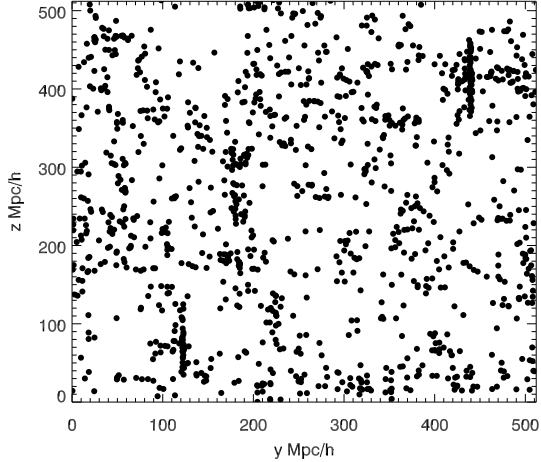,width=3.25in}}
  %\plotone{catalogslice512_pub.ps}
  \caption{We present a 50 \mpch\, slice through our mock catalogue with 
  large 3PCF. Very large "fingers--of--god" are visible e.g. at y 
  $\approx$ z $\approx$ 450 \mpch. } 
  \label{catslice512}
\end{figure}

To understand this further, we show in Figure \ref{catslice512} a 50
\mpch\, thick slice for one of our mock galaxy catalogues generated
using the \it sim1 \rm simulation and using the Group 3 HOD parameters
shown in Figures \ref{allpcf} and \ref{eallpcf}. As expected with such
a high value of $\alpha$ in the HOD, the most massive halos in the
mass function become heavily populated with satellite LRGs and
therefore, produce very strong FOG (along the z-direction). This is clearly not realistic. 

We also note that in the case of Group 3 in Figure \ref{eallpcf}, the
agreement between the real and mock 2PCF has decreased on small
scales.  This is due to the removal of the heavily populated mass
halos in {\it sim1} and {\it sim2}, which mostly affect the 1-halo
term, leaving the 2-halo term (on quasi-linear to linear scales)
unaffected because it is fixed by the initial number density
constraint set on these mock catalogues, i.e., the number density
constraint imposes a mass threshold for the halos which contain a
central galaxy, which in turn, sets the threshold for the mass peaks
that can enter into the 2--point calculation.

\begin{figure}
\centerline{\psfig{file=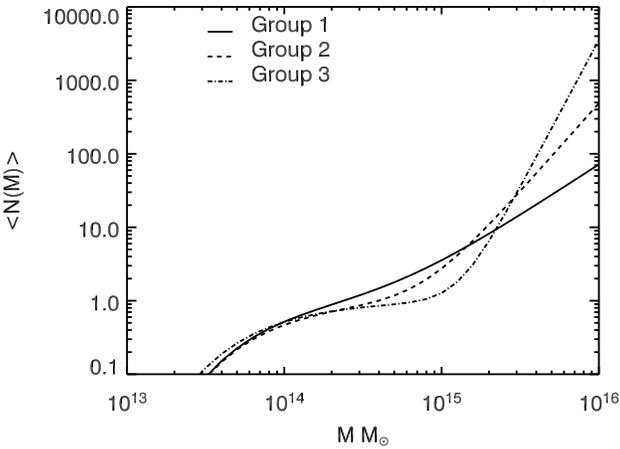,width=3.25in}}
  %\plotone{hodplot_pub.ps}
  \caption{We show here the three HODs ($\left < N(M) \right >$) with
  the lowest 3 $\chi^2$ values shown as black dots in Figure
  \ref{allpcf}.}
  \label{hodnm}
\end{figure}

This work demonstrates the degeneracy between the $M_1$ and $\alpha$
HOD parameters and how one can overpopulate the most massive halos to
compensate for an increase in the $M_1$ parameter, the mass threshold
for adding satellite galaxies to halos. This is clearly illustrated in
the three best--fit parameterized HODs presented in Figure
\ref{hodnm}. All three HOD models have approximately the same minimum
$\chi^2$'s (in Figure \ref{chi2alpha}), and approximately the same
$M_{min}$ (because of the constraint on the number density). Clearly,
as one increases $M_1$, the slope of the HOD must increase to
compensate.

In summary, this work shows that one needs both the 2PCF and 3PCF to
break degeneracies in the HOD parameters, especially $M_1$ and
$\alpha$. We also show in Figure \ref{allpcf} that our 3PCF analysis
favors HOD models with low values of $\alpha$, i.e., it prefers to
populate lower mass halos with satellites at the expense of
overpopulating massive, cluster--like halos with many satellites. This
result agrees with Collister \& Lahav (2005), who find $\alpha\sim1$
for the HOD of red galaxies in clusters and groups of galaxies
detected in the 2dFGRS (see also Popesso et al. 2006). 

This work provides an important insight into the LRG population and
indicates that LRGs are optimal tracers of massive dark matter halos
as there is at least one LRG per halo all the way down to
$M_{halo}\simeq10^{14} M_{\odot}$ (see Figure \ref{hodnm}). Therefore, by
combining the mean observed number density, 2PCF and 3PCF, and using
the halo model to understand how LRGs populate massive halos, one
should be able to accurately test the underlying cosmological
parameters and assumptions of gaussianity. For example, one can test a
suite of different cosmological simulations (by varying parameters
like $\sigma_8$, $\Omega_m$ etc.) against the data, and using the 3PCF
to constrain the HOD parameters, while the large--scale 2PCF would be
sensitive to the changes in the cosmology \citep{Zheng05}. It is
therefore re-assuring that our present best--fit mock catalogue (shown
in the top row of Figure \ref{allpcf}) provides an excellent fit to
all the data presented in this paper, as it is derived from numerical
simulations based on the WMAP1 cosmological parameters (Spergel et
al. 2003) and assumed gaussian initial conditions. Clearly as the data
and mock catalogues improve, we can investigate these issues further
and will explore them in future papers.

\begin{table*}
\caption{The best-fit HOD parameters for the three groups in Figure
\ref{chi2alpha}. Masses are given in $M_{\odot}$. SF and $\rho$ are the 
average satellite fraction and average number density across the mocks
from the six simulations respectively, while $\sigma_{SF}$ and
$\sigma_{\rho}$ are the respective errors. The numbers in the bracket are
the averages across the mocks exlcuding the rogue simulations. }
\begin{tabular}{c|ccccccc}\hline

Group & $M_{min}$ & $M_1$ & $\alpha$ & SF & $\sigma_{SF}$ & $\rho$ & $\sigma_{\rho}$\\ \hline
 1 & 7.66 $\times 10^{13}$ & 4.7 $\times 10^{14}$  &  1.4  & 17.4 (17.3)
 \% & 0.70 (0.21) & $1.02 \times 10^{-4} (1.01 \times 10^{-4}) $ & 0.014
 (0.005) \\ 
 2 & 7.66 $\times 10^{13}$ & 7.6 $\times 10^{14}$  &  2.4  & 10.0 (9.23) 
 \% & 1.64 (0.37) & $9.34 \times 10^{-5} (9.19 \times 10^{-5}) $ & 0.023
 (0.006) \\ 
 3 & 6.66 $\times 10^{13}$ & 1.3 $\times 10^{15}$  &  4.0  & 4.63 (3.24)
 \% & 2.49 (0.26) & $1.03 \times 10^{-4} (1.00 \times 10^{-4}) $ & 0.038
 (0.005) \\ \hline
\end{tabular}
\label{HODtable}
\end{table*}

Finally, in this paper we have focused on the halo model
interpretation of the higher--order LRG clustering rather than the
more established ``biasing'' model for relating dark matter to
galaxies (see Verde et al. 2002, Gaztanaga et al. 2005, Marin et
al. 2006). We plan to explore galaxy biasing in detail in a separate
paper (e.g. Nishimichi et al. 2006) using the LRG 3PCF presented in this paper and dark matter
simulations. However, we can obtain some insight into linear biasing (even in redshift--space, see Nishimichi et al. 2006)
($b$) using the simulations published by GS05. For example, the top
right--hand panel of Figure 2 of GS05 shows that the redshift--space
\Qz for dark matter is close to unity for $30^{\circ} < \theta <
150^{\circ}$ (their $\Lambda$CDM 400 simulation is close to the
simulations used herein and our assumed cosmology). Therefore, we fit
the observed \Qz for the $s=7$ \mpch scale with a constant over the
same range of $\theta$ values and obtain $0.55\pm0.04$. This value is
relatively insensitive to the exact $\theta$ range fitted and
demonstrates that a constant is a good approximation to the form of
\Qz within this $\theta$--range. Assuming linear bias, then $b\simeq
Q_z^{DM}/Q_z^{observed}$, which gives $b=1.83 \pm 0.07$ assuming
$Q_z^{DM}=1$. This confirms our expectation that LRGs are heavily
biased (with respect the the dark matter) and our measured linear bias
value is in excellent agreement with the analysis of Zehavi et
al. (2005), who measured $b=1.84\pm0.11$ (on scales $1<r_p<10$ \mpch)
from the projected 2PCF of the same LRG sample.

\section{Conclusions}
\label{sec:concl}

We present in this paper new measurements of the redshift-space
three-point correlation function of LRGs from the Sloan Digital Sky
Survey (see Figure 2). We have used the same sample as presented in
Eisenstein et al. (2005) and Zehavi et al. (2005) who studied the
two--point correlations function of these LRGs. 
The major conclusions of our work are:

\begin{itemize}

\item We see strong evidence for the expected U--shaped anisotropy in
the shape--dependence of the reduced 3PCF, \Qz, as a function of
$\theta$ (the angle between two sides of the triangle). The evidence
is weakest on the smallest scales probed here ($s=4$ \mpch), where 
\Qz is close to a constant as a function of both $q$ and $\theta$.  We therefore we are not seeing strong "Fingers--of--God" for LRGs as few LRGs are satellite galaxies. On larger scales, a U--shaped anisotropy is predicted (from perturbation theory) and we see evidence for such large--scale structures in our LRG sample.

\item We use jack--knife re--sampling to measure the errors on $Q_z$
and find these errors are stable on all scales to which we are sensitive. We
find that no single jack--knife region dominates the errors, in contrast   
to previous measurements of the 3PCF (Croton et al. 2004, Nichol et
al. 2006). On small scales ($s=4$ and $7$ \mpch), our jack-knife
errors are equal to the expected Poisson errors (based on the number
of triplets in each bin), while for the $s=10$ \mpch scale, our errors
are approximately three times larger than Poisson, indicating they are
correlated due to large--scale structures.

\item We interpret the observed \Qz using a suite of mock galaxy
catalogues generated from large N--body simulations and populated 
with galaxies using the ``halo model'' approach, i.e., we use a
parameterized Halo Occupation Distribution (HOD) to assign LRGs to the
dark matter halos. We find that the combination of the observed number
density of LRGs, the (redshift--space) two--point correlation function
and \Qz provides a strong constraint on the allowed HOD parameters
($M_{min}$, $M_1$, $\alpha$) and breaks key degeneracies between 
  $M_1$ and $\alpha$, the mass threshold for adding a satellite LRG and the slope of the power law for the satellite fraction (Eqns 2 \& 3).

\item The best--fit mock galaxy catalogue to all the data presented in
this paper has HOD parameters of $M_{min} = 7.66 \times 10^{13}
\msun$, $M_1 = 4.7 \times 10^{14}$ and $\alpha=1.4$ (see Eqns 1 \&
2, and Figure 10). Furthermore, we find our \Qz strongly rejects
HODs with higher values of $\alpha$ as they overpopulate the massive
halos leading to stronger ``Fingers--of--God'' than witnessed in the
real data (see Gazta{\~na}ga \& Scoccimarro 2005).  As shown in Table \ref{HODtable}, only 17\% of LRGs are satellite galaxies. 

\item Assuming linear biasing between the dark matter and LRGs on the
scales probed here, we estimate $b=1.83 \pm 0.07$ (assuming
$Q_z^{DM}=1)$. 
This value is in excellent agreement with Zehavi et
al. (2005) based on the projected 2PCF of the same LRG sample.

\end{itemize}
To facilitate further analysis of the LRG sample, the 2PCF, 3PCF, mass functions and covariance matrices presented in this paper are available on request from Bob Nichol and/or via the website {\tt http://www.dsg.port.ac.uk/\~\,nicholb/3pt/kulkarni/ }. 

\section*{Acknowledgments}

We thank the referee for the comments. We thank Rupert Croft, Peder Norberg, Kathy Romer, David Weinberg, Larry Wasserman and Zheng Zheng for stimulating conversations and important insights about our work.
We thank Quentin Mercer III, Rupert Croft and Albert Wong for their
help and assistance in building and maintainace of astrophysics
Beowulf Cluster at Carnegie Mellon University. We also thank the
administrators of TeraGrid at NCSA for their help in running 3PCF on
the NCSA supercomputer.  The work presented here was partially funded
by NSF ITR Grant 0121671.  RN also thanks the EU Marie Curie program
for partial funding during this work.

Funding for the SDSS and SDSS-II has been provided by the Alfred P. Sloan Foundation, the Participating Institutions, the National Science Foundation, the U.S. Department of Energy, the National Aeronautics and Space Administration, the Japanese Monbukagakusho, the Max Planck Society, and the Higher Education Funding Council for England. The SDSS Web Site is http://www.sdss.org/.

The SDSS is managed by the Astrophysical Research Consortium for the Participating Institutions. The Participating Institutions are the American Museum of Natural History, Astrophysical Institute Potsdam, University of Basel, Cambridge University, Case Western Reserve University, University of Chicago, Drexel University, Fermilab, the Institute for Advanced Study, the Japan Participation Group, Johns Hopkins University, the Joint Institute for Nuclear Astrophysics, the Kavli Institute for Particle Astrophysics and Cosmology, the Korean Scientist Group, the Chinese Academy of Sciences (LAMOST), Los Alamos National Laboratory, the Max-Planck-Institute for Astronomy (MPIA), the Max-Planck-Institute for Astrophysics (MPA), New Mexico State University, Ohio State University, University of Pittsburgh, University of Portsmouth, Princeton University, the United States Naval Observatory, and the University of Washington.

\appendix

\section{The 3PCF Data}
\label{3pcf_data}

Following tables contain the values of the data points in the figure
\ref{lrg3pcf}. The errors given here are our estimates from jack-knife
resampling and represent the diagonal elements of the covariance matrix
The column named `DDD' gives the number of triplets in the data for the
corresponding bin.

\clearpage
%\LongTables
\begin{landscape}
\begin{table}
\caption{3PCF measurements for $3.7 s < 4.3 $\mpch}
\begin{tabular}{cc|cccccccccccccccccc}
$\theta_{min}$ & $\theta_{max}$ &    & DDD & \zetar & $\sigma_{\zeta}$ & \Qz & $\sigma_{Q_z}$ &    & DDD & \zetar & $\sigma_{\zeta}$ &\Qz & $\sigma_{Q_z}$ &    & DDD & \zetar & $\sigma_{\zeta}$ &
\Qz & $\sigma_{Q_z}$ \\ \hline
 & & \multicolumn{5}{c}{$0.9 < q < 1.1$} &  & \multicolumn{5}{c}{$1.9 <
 q < 2.1$}  &  & \multicolumn{5}{c}{$2.9 < q < 3.1$} \\
 \cline{4-8} \cline{10-14}  \cline{16-20} \\ \hline
0.020  &0.040  &  & 19   &426.397 & 822.366 & 0.707 & 0.264 &  & 172   &65.937 & 8.766 & 0.835 & 0.117 &  &  417   &21.340 & 2.246 & 0.841 & 0.090 \\
0.060  &0.080  &  &  37  &350.639 &  74.754 & 0.748 & 0.118 &   & 201
&57.863 & 8.052 & 0.762 & 0.111 &  &   460  &20.422 & 2.188 & 0.826 &
0.091  \\
0.100  &0.120  &  &  57  &244.196 &  65.802 & 0.755 & 0.168 &   & 262
&54.209 & 6.128 & 0.770 & 0.094 &  &   509  &17.955 & 1.948 & 0.753 &
0.078  \\
0.140  &0.160  &  &  61  &157.611 &  30.985 & 0.576 & 0.098 &   & 326
&46.236 & 4.215 & 0.722 & 0.072 &  &   613  &17.547 & 1.629 & 0.771 &
0.063  \\
0.180  &0.200  &  &  59  &106.098 &  23.236 & 0.467 & 0.096 &   & 368
&39.005 & 3.510 & 0.670 & 0.062 &  &   687  &15.411 & 1.730 & 0.710 &
0.058  \\
0.220  &0.240  &  &  81  &111.398 &  12.279 & 0.583 & 0.044 &   & 372
&31.463 & 3.510 & 0.598 & 0.059 &  &   777  &14.387 & 1.362 & 0.704 &
0.048  \\
0.260  &0.280  &  &  92  &109.073 &   9.469 & 0.640 & 0.043 &   & 376
&27.098 & 3.910 & 0.569 & 0.074 &  &   839  &13.126 & 1.161 & 0.679 &
0.040  \\
0.300  &0.320  &  &  67  &101.938 &  13.075 & 0.664 & 0.081 &   & 362
&23.166 & 2.796 & 0.527 & 0.057 &  &   832  &11.734 & 1.176 & 0.634 &
0.040  \\
0.340  &0.360  &  &  66  & 83.169 &   5.983 & 0.601 & 0.043 &   & 331
&20.863 & 2.423 & 0.511 & 0.046 &  &   785  &10.943 & 1.255 & 0.618 &
0.050  \\
0.380  &0.400  &  & 110  & 76.032 &   8.263 & 0.604 & 0.060 &   & 269
&19.024 & 2.600 & 0.497 & 0.059 &  &   683  &10.023 & 1.096 & 0.595 &
0.044  \\
0.420  &0.440  &  & 135  & 65.968 &   7.125 & 0.569 & 0.060 &   & 240
&17.607 & 3.003 & 0.493 & 0.074 &  &   541  & 9.116 & 1.216 & 0.564 &
0.058  \\
0.460  &0.480  &  & 155  & 62.264 &   9.201 & 0.573 & 0.081 &   & 349
&17.588 & 2.720 & 0.524 & 0.067 &  &   611  & 9.063 & 1.349 & 0.583 &
0.067  \\
0.500  &0.520  &  & 157  & 55.097 &   7.701 & 0.538 & 0.061 &   & 432
&17.541 & 2.915 & 0.552 & 0.076 &  &   791  & 8.986 & 1.331 & 0.599 &
0.070  \\
0.540  &0.560  &  & 166  & 53.064 &   8.110 & 0.548 & 0.064 &   & 525
&18.929 & 3.170 & 0.624 & 0.086 &  &   941  & 8.833 & 1.213 & 0.611 &
0.066  \\
0.580  &0.600  &  & 203  & 62.454 &   6.711 & 0.670 & 0.053 &   & 573
&18.740 & 2.889 & 0.639 & 0.083 &  &  1057  & 8.887 & 1.045 & 0.636 &
0.057  \\
0.620  &0.640  &  & 214  & 62.468 &   6.236 & 0.692 & 0.057 &   & 619
&19.918 & 2.256 & 0.697 & 0.068 &  &  1128  & 9.095 & 0.926 & 0.669 &
0.048  \\
0.660  &0.680  &  & 222  & 65.282 &   6.752 & 0.745 & 0.072 &   & 623
&20.388 & 2.819 & 0.727 & 0.091 &  &  1126  & 8.981 & 0.809 & 0.679 &
0.043  \\
0.700  &0.720  &  & 214  & 63.369 &   7.396 & 0.738 & 0.076 &   & 604  &20.587 & 2.497 & 0.752 & 0.081 &  &  1105  & 9.184 & 0.868 & 0.706 & 0.047 \\
0.740  &0.760  &  & 204  & 62.464 &   7.932 & 0.743 & 0.088 &   & 572
&20.626 & 2.295 & 0.768 & 0.074 &  &  1040  & 9.115 & 0.905 & 0.715 &
0.053  \\
0.780  &0.800  &  & 194  & 62.797 &   9.055 & 0.756 & 0.105 &   & 552
&21.731 & 2.263 & 0.823 & 0.078 &  &   971  & 9.186 & 0.785 & 0.734 &
0.045  \\
0.820  &0.840  &  & 181  & 63.600 &   9.996 & 0.778 & 0.122 &   & 513
&22.037 & 2.181 & 0.843 & 0.078 &  &   901  & 9.205 & 0.596 & 0.743 &
0.037  \\
0.860  &0.880  &  & 174  & 65.524 &  10.185 & 0.809 & 0.127 &   & 466
&21.649 & 2.280 & 0.836 & 0.081 &  &   831  & 9.166 & 0.614 & 0.750 &
0.042  \\
0.900  &0.920  &  & 163  & 65.106 &   9.223 & 0.813 & 0.120 &   & 436
&21.718 & 2.307 & 0.844 & 0.088 &  &   780  & 9.236 & 0.685 & 0.761 &
0.048  \\
0.940  &0.960  &  & 158  & 66.026 &   9.590 & 0.828 & 0.127 &   & 411
&21.396 & 2.308 & 0.836 & 0.091 &  &   756  & 9.490 & 0.729 & 0.784 &
0.053  \\
0.980  &1.000  &  & 155  & 66.075 &  10.032 & 0.829 & 0.132 &   & 405
&21.764 & 2.304 & 0.851 & 0.092 &  &   747  & 9.695 & 0.776 & 0.802 &
0.057  \\ \hline
\end{tabular}
\end{table}

\clearpage
\begin{table}

%\centering
\caption{3PCF measurements for $6.9 s < 7.1 $\mpch}

\begin{tabular}{cc|cccccccccccccccccc}

$\theta_{min}$ & $\theta_{max}$ &    &
DDD & \zetar & $\sigma_{\zeta}$ &
\Qz & $\sigma_{Q_z}$ &    &
DDD & \zetar & $\sigma_{\zeta}$ &
\Qz & $\sigma_{Q_z}$ &    &
DDD & \zetar & $\sigma_{\zeta}$ &
\Qz & $\sigma_{Q_z}$  \\  \hline

 & & \multicolumn{5}{c}{$0.9 < q < 1.1$} &  & \multicolumn{5}{c}{$1.9 <
 q < 2.1$}  &  & \multicolumn{5}{c}{$2.9 < q < 3.1$} \\
 \cline{4-8} \cline{10-14}  \cline{16-20} \\ \hline
 
0.020   &0.040   &   &  16   &102.117   &27.123   &0.589   &0.170  &   &  92   &13.319  &2.875   &1.097   &0.263   &   &  92   &4.767  &0.936  &1.401 & 0.326  \\  
0.060   &0.080   &   &  26   & 59.691   &21.345   &0.538   &0.182  &   & 106   &10.485  &1.878   &0.909   &0.180   &   & 106   &4.352  &0.865  &1.305 & 0.285  \\  
0.100   &0.120   &   &  35   & 46.977   &14.803   &0.601   &0.179  &   & 121   & 7.680  &1.334   &0.714   &0.138   &   & 121   &3.546  &0.858  &1.103 & 0.274  \\  
0.140   &0.160   &   &  42   & 32.994   & 6.888   &0.583   &0.113  &   & 128   & 5.363  &0.752   &0.563   &0.090   &   & 128   &2.817  &0.581  &0.922 & 0.197  \\  
0.180   &0.200   &   &  42   & 22.151   & 4.311   &0.495   &0.097  &   & 133   & 4.584  &1.028   &0.546   &0.103   &   & 133   &2.203  &0.500  &0.761 & 0.141  \\  
0.220   &0.240   &   &  57   & 24.984   & 4.697   &0.707   &0.121  &   & 132   & 3.657  &0.820   &0.488   &0.104   &   & 132   &2.054  &0.407  &0.752 & 0.119  \\  
0.260   &0.280   &   &  64   & 22.355   & 5.427   &0.755   &0.168  &   & 121   & 2.066  &0.637   &0.301   &0.096   &   & 121   &1.782  &0.571  &0.681 & 0.177  \\  
0.300   &0.320   &   &  44   & 16.826   & 6.547   &0.672   &0.250  &   & 132   & 2.265  &0.652   &0.365   &0.117   &   & 132   &1.604  &0.515  &0.648 & 0.179  \\  
0.340   &0.360   &   &  36   &  9.797   & 3.850   &0.435   &0.164  &   & 142   & 2.802  &0.907   &0.491   &0.159   &   & 142   &1.343  &0.491  &0.572 & 0.188  \\  
0.380   &0.400   &   &  59   &  9.263   & 2.161   &0.452   &0.103  &   & 125   & 3.130  &0.847   &0.590   &0.161   &   & 125   &0.865  &0.350  &0.391 & 0.149  \\  
0.420   &0.440   &   &  67   &  9.646   & 1.771   &0.531   &0.090  &   & 100   & 3.045  &0.760   &0.625   &0.149   &   & 100   &0.642  &0.380  &0.306 & 0.181  \\  
0.460   &0.480   &   &  73   &  9.094   & 2.244   &0.545   &0.120  &   & 165   & 3.306  &0.930   &0.725   &0.188   &   & 165   &0.525  &0.336  &0.264 & 0.167  \\  
0.500   &0.520   &   &  69   &  5.851   & 1.552   &0.366   &0.096  &   & 174   & 2.669  &0.950   &0.614   &0.196   &   & 174   &0.654  &0.234  &0.347 & 0.114  \\  
0.540   &0.560   &   &  82   &  7.607   & 1.217   &0.496   &0.082  &   & 160   & 1.743  &0.685   &0.420   &0.153   &   & 160   &0.970  &0.325  &0.546 & 0.173  \\  
0.580   &0.600   &   &  83   &  7.101   & 1.528   &0.486   &0.118  &   & 177   & 2.225  &0.673   &0.555   &0.150   &   & 177   &1.085  &0.435  &0.630 & 0.220  \\  
0.620   &0.640   &   &  96   &  8.468   & 1.907   &0.602   &0.152  &   & 185   & 2.140  &0.629   &0.549   &0.153   &   & 185   &1.135  &0.458  &0.676 & 0.217  \\  
0.660   &0.680   &   &  91   &  6.967   & 1.670   &0.510   &0.132  &   & 211   & 2.857  &0.686   &0.745   &0.177   &   & 211   &1.013  &0.431  &0.615 & 0.217  \\  
0.700   &0.720   &   & 112   &  9.696   & 1.912   &0.729   &0.169  &   & 236   & 3.570  &0.931   &0.962   &0.238   &   & 236   &1.095  &0.304  &0.682 & 0.157  \\  
0.740   &0.760   &   & 121   & 11.390   & 2.085   &0.871   &0.183  &   & 232   & 3.577  &0.714   &0.985   &0.210   &   & 232   &1.247  &0.339  &0.787 & 0.197  \\  
0.780   &0.800   &   & 112   & 11.276   & 1.913   &0.877   &0.173  &   & 231   & 4.198  &0.791   &1.180   &0.247   &   & 231   &1.328  &0.226  &0.850 & 0.126  \\  
0.820   &0.840   &   & 105   & 11.677   & 2.097   &0.920   &0.191  &   & 210   & 4.418  &0.998   &1.258   &0.319   &   & 210   &1.552  &0.308  &1.015 & 0.213  \\  
0.860   &0.880   &   &  95   & 11.574   & 2.398   &0.923   &0.212  &   & 184   & 4.519  &1.055   &1.307   &0.337   &   & 184   &1.720  &0.404  &1.115 & 0.263  \\  
0.900   &0.920   &   &  92   & 13.354   & 2.652   &1.079   &0.237  &   & 156   & 4.387  &0.899   &1.270   &0.296   &   & 156   &1.800  &0.358  &1.175 & 0.221  \\  
0.940   &0.960   &   &  89   & 14.399   & 3.144   &1.170   &0.280  &   & 143   & 4.734  &0.972   &1.380   &0.328   &   & 143   &1.906  &0.450  &1.245 & 0.281  \\  
0.980   &1.000   &   &  84   & 14.289   & 3.163   &1.165   &0.284  &   & 135   & 4.783  &1.031   &1.392   &0.344   &   & 135   &1.934  &0.387  &1.267 & 0.247  \\  \hline
\end{tabular}
\end{table}

\clearpage
\begin{table}

%\centering
\caption{3PCF measurements for $9.9 s < 10.1 $\mpch}
\begin{tabular}{cccccccccccccccccccc}

$\theta_{min}$ & $\theta_{max}$ &    &
DDD & \zetar & $\sigma_{\zeta}$ &
\Qz & $\sigma_{Q_z}$ &    &
DDD & \zetar & $\sigma_{\zeta}$ &
\Qz & $\sigma_{Q_z}$ &    &
DDD & \zetar & $\sigma_{\zeta}$ &
\Qz & $\sigma_{Q_z}$  \\  \hline
 & & \multicolumn{5}{c}{$0.9 < q < 1.1$} &  & \multicolumn{5}{c}{$1.9 <
 q < 2.1$}  &  & \multicolumn{5}{c}{$2.9 < q < 3.1$} \\
 \cline{4-8} \cline{10-14}  \cline{16-20} \\ \hline
   
0.020  &0.040  &  &  40  &35.077  &9.336 & 0.565  &0.115 &   & 183  &3.802 & 0.736 & 1.167  &0.221 &   & 220  &0.999 & 0.297  &1.203  &0.354   \\
0.060  &0.080  &  &  62  &21.049  &4.319 & 0.536  &0.102 &   & 228  &3.087 & 0.627 & 0.993  &0.191 &   & 279  &0.983 & 0.230  &1.223  &0.289   \\
0.100  &0.120  &  &  95  &19.423  &2.093 & 0.753  &0.068 &   & 281  &2.286 & 0.359 & 0.804  &0.126 &   & 362  &0.882 & 0.176  &1.140  &0.240   \\
0.140  &0.160  &  &  82  & 8.381  &2.189 & 0.476  &0.119 &   & 312  &1.526 & 0.357 & 0.596  &0.142 &   & 422  &0.657 & 0.157  &0.898  &0.228   \\
0.180  &0.200  &  &  98  & 7.727  &1.858 & 0.600  &0.142 &   & 308  &1.089 & 0.420 & 0.480  &0.191 &   & 469  &0.524 & 0.136  &0.785  &0.238   \\
0.220  &0.240  &  & 103  & 5.876  &1.161 & 0.585  &0.116 &   & 318  &1.127 & 0.513 & 0.566  &0.260 &   & 458  &0.385 & 0.137  &0.633  &0.250   \\
0.260  &0.280  &  &  96  & 3.095  &1.266 & 0.373  &0.149 &   & 334  &0.952 & 0.419 & 0.538  &0.243 &   & 485  &0.440 & 0.097  &0.777  &0.219   \\
0.300  &0.320  &  &  64  & 1.132  &0.921 & 0.168  &0.133 &   & 330  &0.663 & 0.229 & 0.414  &0.148 &   & 513  &0.488 & 0.107  &0.897  &0.205   \\
0.340  &0.360  &  &  85  & 2.739  &0.841 & 0.449  &0.127 &   & 310  &0.382 & 0.186 & 0.260  &0.123 &   & 503  &0.315 & 0.112  &0.624  &0.218   \\
0.380  &0.400  &  & 125  & 2.331  &0.692 & 0.426  &0.124 &   & 259  &0.351 & 0.370 & 0.256  &0.271 &   & 467  &0.311 & 0.175  &0.648  &0.348   \\
0.420  &0.440  &  & 147  & 2.593  &0.775 & 0.518  &0.158 &   & 228  &0.573 & 0.508 & 0.455  &0.411 &   & 361  &0.426 & 0.226  &0.928  &0.444   \\
0.460  &0.480  &  & 151  & 1.855  &0.549 & 0.398  &0.117 &   & 323  &0.168 & 0.360 & 0.143  &0.310 &   & 416  &0.359 & 0.183  &0.840  &0.364   \\
0.500  &0.520  &  & 166  & 1.702  &0.746 & 0.397  &0.177 &   & 387  &0.360 & 0.229 & 0.333  &0.218 &   & 578  &0.417 & 0.156  &1.006  &0.295   \\
0.540  &0.560  &  & 164  & 1.100  &0.800 & 0.271  &0.200 &   & 423  &0.546 & 0.157 & 0.535  &0.151 &   & 610  &0.332 & 0.134  &0.829  &0.305   \\
0.580  &0.600  &  & 194  & 1.881  &0.569 & 0.487  &0.152 &   & 489  &0.939 & 0.166 & 0.950  &0.158 &   & 634  &0.332 & 0.118  &0.864  &0.272   \\
0.620  &0.640  &  & 201  & 1.672  &0.647 & 0.449  &0.179 &   & 506  &0.963 & 0.244 & 1.007  &0.245 &   & 622  &0.241 & 0.101  &0.659  &0.239   \\
0.660  &0.680  &  & 220  & 2.173  &0.579 & 0.597  &0.164 &   & 528  &1.109 & 0.221 & 1.193  &0.269 &   & 639  &0.225 & 0.125  &0.642  &0.320   \\
0.700  &0.720  &  & 248  & 2.691  &0.668 & 0.758  &0.188 &   & 548  &1.143 & 0.248 & 1.254  &0.307 &   & 672  &0.346 & 0.158  &0.985  &0.385   \\
0.740  &0.760  &  & 246  & 2.728  &0.592 & 0.782  &0.164 &   & 528  &1.033 & 0.327 & 1.147  &0.381 &   & 694  &0.460 & 0.171  &1.374  &0.429   \\
0.780  &0.800  &  & 239  & 3.123  &0.647 & 0.912  &0.188 &   & 492  &1.051 & 0.326 & 1.201  &0.390 &   & 650  &0.443 & 0.177  &1.355  &0.464   \\
0.820  &0.840  &  & 230  & 3.508  &0.730 & 1.040  &0.211 &   & 427  &0.904 & 0.307 & 1.046  &0.355 &   & 598  &0.547 & 0.221  &1.683  &0.585   \\
0.860  &0.880  &  & 205  & 3.482  &0.724 & 1.045  &0.210 &   & 388  &1.078 & 0.320 & 1.259  &0.381 &   & 524  &0.597 & 0.245  &1.831  &0.651   \\
0.900  &0.920  &  & 185  & 3.589  &0.763 & 1.088  &0.224 &   & 330  &1.031 & 0.247 & 1.208  &0.291 &   & 451  &0.630 & 0.239  &1.939  &0.660   \\
0.940  &0.960  &  & 168  & 3.669  &0.803 & 1.117  &0.235 &   & 294  &1.142 & 0.275 & 1.356  &0.314 &   & 391  &0.595 & 0.256  &1.852  &0.750   \\
0.980  &1.000  &  & 162  & 3.794  &0.879 & 1.157  &0.257 &   & 265  &1.002 & 0.339 & 1.198  &0.402 &   & 361  &0.588 & 0.260  &1.834  &0.755   \\ \hline
\end{tabular}
\end{table}

\clearpage
\end{landscape}
%\end{appendix}

\end{document}